\documentclass[aps, prx, twocolumn,longbibliography,superscriptaddress,amsmath,amssymb,floatfix]{revtex4-2}

\usepackage[version=4]{mhchem}
\usepackage{amssymb}
\usepackage{graphicx}
\usepackage{amsmath}
\usepackage{color}
\usepackage{mathrsfs}
\usepackage{float}
\usepackage{indentfirst}
\usepackage{mathrsfs}
\usepackage{float}
\usepackage{indentfirst}
\usepackage{textcomp}
\usepackage{comment}
\usepackage{mathtools}
\usepackage{natbib,hyperref}
\usepackage{soul}  
\usepackage{bm}

\begin{document}

\title{Augmenting Density Matrix Renormalization Group with Matchgates and Clifford circuits}

\author{Jiale Huang}
\altaffiliation{These authors contributed equally to this work.}
\affiliation{Key Laboratory of Artificial Structures and Quantum Control (Ministry of Education),  School of Physics and Astronomy, Shanghai Jiao Tong University, Shanghai 200240, China}

\author{Xiangjian Qian}
\altaffiliation{These authors contributed equally to this work.}
\affiliation{Key Laboratory of Artificial Structures and Quantum Control (Ministry of Education),  School of Physics and Astronomy, Shanghai Jiao Tong University, Shanghai 200240, China}
\affiliation{Tsung-Dao Lee Institute, Shanghai Jiao Tong University, Shanghai 200240, China}

\author{Zhendong Li} 
\thanks{zhendongli@bnu.edu.cn}
\affiliation{Key Laboratory of Theoretical and Computational Photochemistry, Ministry of Education, College of Chemistry, Beijing Normal University, Beijing 100875, China}

\author{Mingpu Qin} \thanks{qinmingpu@sjtu.edu.cn}
\affiliation{Key Laboratory of Artificial Structures and Quantum Control (Ministry of Education),  School of Physics and Astronomy, Shanghai Jiao Tong University, Shanghai 200240, China}

\affiliation{Hefei National Laboratory, Hefei 230088, China}

\date{\today}


\begin{abstract}
Matchgates and Clifford circuits are two types of quantum circuits which can be efficiently simulated classically, though the underlying reasons are quite different. Matchgates are essentially the single particle basis transformations in the Majorana fermion representation, which can be easily handled classically, while the Clifford circuits can be efficiently simulated using the tableau method according to the Gottesman-Knill theorem. In this work, we propose a new wave-function ansatz in which matrix product states are augmented with the combination of Matchgates and Clifford circuits (dubbed MCA-MPS) to take advantage of the representing power of all of them. Moreover, the optimization of MCA-MPS can be efficiently implemented within the Density Matrix Renormalization Group method. Our benchmark results on one-dimensional hydrogen chain show that MCA-MPS can improve the accuracy of the ground-state calculation by several orders of magnitude over MPS with the same bond dimension. This new method provides us a useful approach to study quantum many-body systems. The MCA-MPS ansatz also expands our understanding of classically simulatable quantum many-body states. 
\end{abstract}

\maketitle
{\em Introduction--}
The classical simulation of strongly correlated quantum many-body systems is one of the most important tasks in the study of condensed matter physics. Over the past decades, many different types of many-body methods were proposed \cite{RevModPhys.68.13,RevModPhys.78.865,RevModPhys.87.1067,RevModPhys.73.33,RevModPhys.93.045003,xiang2023density,PhysRevX.5.041041} which can effectively handle different categories of quantum systems. Tensor Network State (TNS) is one representative of them which stems from the famous Density Matrix Renormalization Group (DMRG) method \cite{PhysRevLett.69.2863}. DMRG is now the workhorse for one dimensional (1D) quantum many-body systems because the underlying wave-function ansatz, Matrix Product States (MPS) \cite{PhysRevLett.75.3537}, can capture the entanglement structure of 1D quantum systems. DMRG has difficulty for higher dimensional quantum systems and new ansatzes were proposed by generalizing MPS to higher dimension (such as PEPS \cite{2004cond.mat..7066V}, PESS \cite{PhysRevX.4.011025}, MERA \cite{PhysRevLett.102.180406,PhysRevLett.99.220405} and so on \cite{RevModPhys.93.045003,xiang2023density}) to encode the entanglement entropy area law \cite{RevModPhys.82.277} of quantum many-body systems. Nevertheless, DMRG is still widely used in the study of quasi-1D systems because of its low cost and accurate ground state can be obtained by pushing the bond dimension to large numbers. There are also attempts to enhance the representing power of MPS by augmenting it with disentanglers with its friendly computational cost retained \cite{Qian_2023}.   

The developments in the field of quantum computing and quantum information \cite{Nielsen_Chuang_2010} have identified special quantum circuits which can be efficiently simulated classically. Clifford circuits \cite{gottesman1997stabilizer} and Matchgates \cite{10.1145/380752.380785} are two types of classically simulatable quantum circuits. In the (Majorana) fermion representation, Matchgates is essentially the evolution of Gaussian states \cite{doi:10.1098/rspa.2008.0189}, which can be easily handled classically. Clifford circuits that consist solely of Clifford gates (Hadamard gate, the phase gate S, and the controlled-NOT gate) can be efficiently simulated using the tableau method according to the Gottesman–Knill theorem \cite{gottesman1997stabilizer,PhysRevA.70.052328,PhysRevA.73.022334}. Giving that MPS, Matchgates, Clifford circuits can efficiently describe different types of quantum states (low-entangled states, stabilizer states, Gaussian states respectively), it is tempting to combine them to construct more powerful ansatz. 

In \cite{PhysRevLett.133.190402}, the Clifford Circuits Augmented Matrix Product States (CAMPS) method was proposed (also see \cite{PhysRevLett.133.230601,PRXQuantum.6.010345}), in which MPS is augmented with Clifford circuits. To solve the ground state of a quantum system with CAMPS, the MPS part needs only to deal with the so called Non-stabilizerness Entanglement Entropy \cite{2024arXiv240916895H} because the Clifford circuits can take care of the contribution from stabilizer, making the accuracy significantly higher than MPS with the same bond dimensions. The optimization of both MPS and Clifford circuits can be efficiently implemented by a slight modification of the DMRG method. Later on, CAMPS was also generalized to the study of time evolution \cite{PhysRevLett.134.150404,PhysRevLett.134.150403} and the calculation of finite temperature properties \cite{2024arXiv241015709Q}. It was also applied to fermion models \cite{2025arXiv250100413H} by mapping the fermion degrees of freedom to spin ones with the Jordan-Wigner transformation. With CAMPS, it was also found that duality (like the Kramer-Wannier self-duality) in certain one-dimensional quantum system can be represented as Clifford circuits \cite{PhysRevB.111.085121,2024arXiv241111720F}. 

\begin{figure*}
    \centering
    \includegraphics[width=\linewidth]{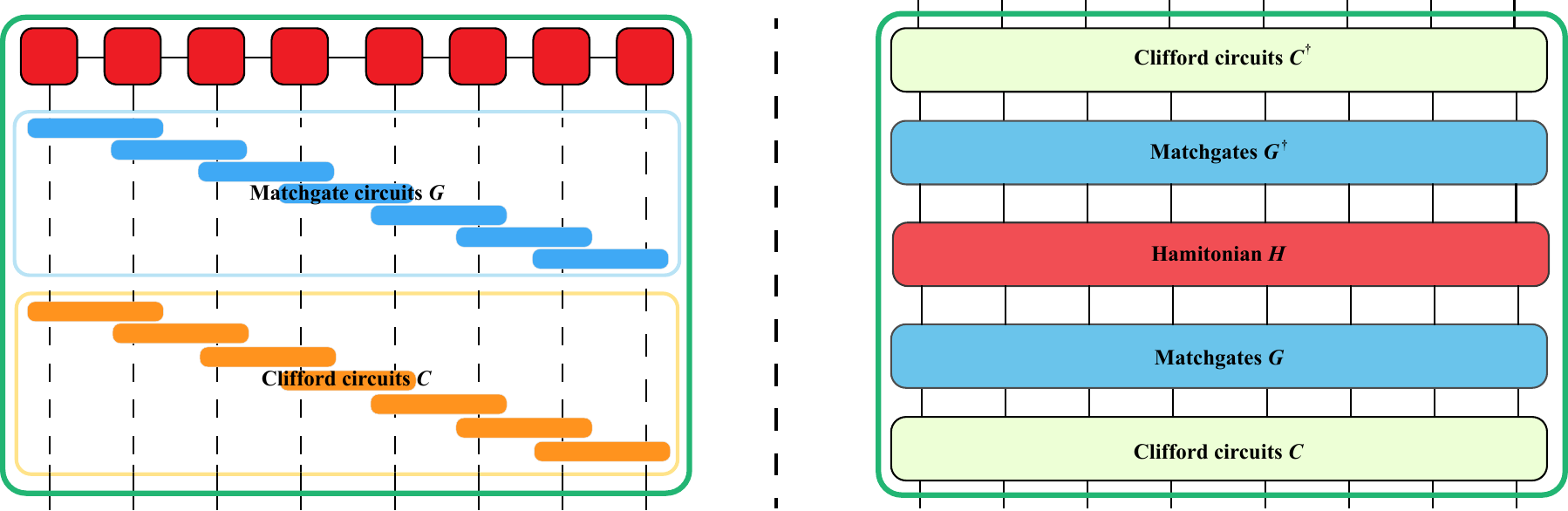}
    \caption{Schematic illustration of MCA-MPS. Left: The structure of the MCA-MPS ansatz, $|\text{MCA-MPS}\rangle = C G |\text{MPS}\rangle$, depicting the sequential application of Matchgate circuits $G$ and Clifford circuits $C$ to the Matrix Product State $|\text{MPS}\rangle$ corresponding to Hamiltonian $H$. Notice that we only deal with $|\text{MCA-MPS}\rangle$ which is less entangled than $|\text{MPS}\rangle$. Right: The equivalent transformation of the Hamiltonian, where the original Hamiltonian $H$ is transformed to a effective Hamiltonian ($H_\text{MCA-MPS} = C G H G^\dagger C^\dagger$) by the successive application of Matchgates and Clifford circuits.}
    \label{MCAMPS}
\end{figure*}

Matchgates were first proposed in \cite{10.1145/380752.380785} and are proved to be classically simulatable. Later on, it was realized that the circuits built from Matchgates are essentially the transformations describing the evolution of Gaussian states \cite{doi:10.1098/rspa.2008.0189}, which is known to be classically simulatable. In this regard, the studies of the combination of Matchgates and MPS has a long history in the form of single particle basis transformation \cite{PhysRevLett.117.210402,PhysRevB.111.035119}.    

In this work, we propose a new ansatz in which MPS is augmented with both Matchgates and Clifford circuits (dubbed MCA-MPS). The representing power of the integration of Matchgates and Clifford circuits was previously studied in \cite{2024arXiv241010068P}. It is known that Clifford circuits can describe quantum states with large entanglement \cite{PhysRevX.7.031016} (and correlation) but they are not universal for quantum computing. Clifford circuits can only describe stabilizer state. The distance between a given quantum state and a stabilizer state is usually called magic or non-stabilizerness and many quantities has been proposed to quantify it \cite{Bravyi2019simulationofquantum,PhysRevX.6.021043,Bravyi2019simulationofquantum,PhysRevLett.116.250501,Veitch_2014,PhysRevLett.115.070501,PhysRevLett.128.050402}. The quantum states describe by Matchgates are Gaussian states, which have no two-body correlation but can host large magic \cite{2024arXiv241205367C}. In this sense, the combination of Clifford circuits and Matchgates can complement each other to increase the representing power. It is known that MPS can describe ``general'' low-entangled states. Given these facts, the MCA-MPS, which take advantage of the representing ability of all ingredients in it, should be a powerful ansatz. We show that the optimization of all the MPS, Matchgates, Clifford circuits in MCA-MPS can be efficiently implemented by a slight modification of the DMRG method. We also benchmark MCA-MPS on one-dimensional hydrogen chain system and two-dimensional hydrogen lattice (see Supplemental Material \cite{supp}). The results show MCA-MPS can significantly improve the accuracy of ground state energy by several orders of magnitude over MPS with the same bond dimension.

{\em Method--} To synergistically leverage the distinct representational capabilities of Matrix Product States, Matchgate, and Clifford circuits, we introduce the new MCA-MPS ansatz. Its structure, depicted in Fig. \ref{MCAMPS}, is defined as:
\begin{equation}
    |\text{MCA-MPS}\rangle = C G |\text{MPS}\rangle
    \label{MCA-MPS ansatz}
\end{equation}
where $|\text{MPS}\rangle$ represents a standard MPS efficiently capturing low entanglement state, $G$ denotes Matchgate circuits, and $C$ are Clifford circuits. Here, the Matchgate and Clifford circuits act as disentanglers applied to the original $|\text{MPS}\rangle$, resulting in the final ansatz $|\text{MCA-MPS}\rangle$. 
Matchgate circuits $G$ implement $SO(2N)$ transformations within the Majorana fermion representation \cite{doi:10.1098/rspa.2008.0189} ($\gamma_i \rightarrow O_{i,j} \gamma_j$ with matrix $O \in SO(2N)$ and $\gamma_i$ are Majorana operators). This component enables the efficient representation of Gaussian states. The subsequent Clifford circuits $C$ further enhances the ansatz by applying Pauli-basis transformations.

The MCA-MPS ansatz is variationally optimized using a modified Density Matrix Renormalization Group algorithm through a sequential procedure. First, the Matrix Product State component is initialized and refined via a standard DMRG optimization to yield a suitable initial state. Subsequently, the Matchgate circuits G are introduced and optimized to reduce entanglement entropy and further minimize the ground-state energy (see Ref. \cite{li2025entanglementminimizedorbitalsenablefaster} for details). Physically, this optimization of Matchgates is equivalent to an adaptive single-particle basis rotation. Following this, the Clifford circuits $C$ is applied and optimized to achieve additional energy and entanglement reduction (see Ref. \cite{PhysRevLett.133.190402} for details). These optimization steps are integrated within the DMRG framework. Both the Matchgates and Clifford circuit optimization can be performed after the eigen-state is obtained for the effective Hamiltonian in DMRG and before the truncation is performed with a singular value decomposition \cite{PhysRevLett.133.190402}. The optimization of Clifford circuits is conducted in the Pauli basis. Matchgate circuit optimization can proceed in either the Pauli or fermionic basis. If the fermionic basis is employed for the Matchgate optimization, a Jordan-Wigner transformation is utilized to map fermionic operators to spin operators prior to the optimization of Clifford circuits.

\begin{figure}[t]
    \centering
    \includegraphics[width=\linewidth]{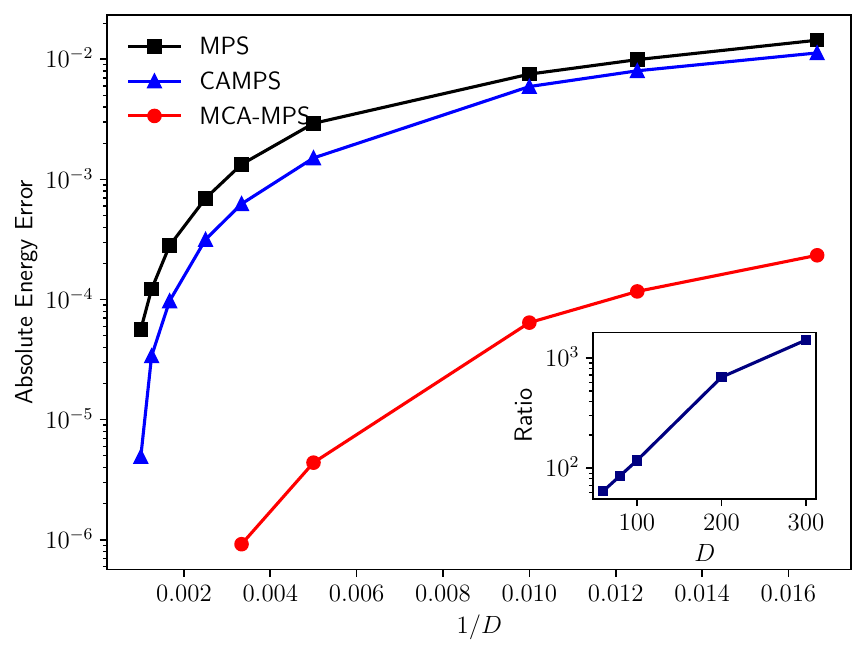}
    \caption{Ground-state energy error (in the unit of Hartree) for the 1D hydrogen chain \ce{H12} (with restricted Hartree-Fock orbital) relative to exact energy, plotted against bond dimension D. Results are shown for both MPS, CAMPS and MCA-MPS. Errors are defined as  $|E_{\text{exact}} - E_{\text{MPS}}|$, $|E_{\text{exact}} - E_{\text{CAMPS}}|$ and $|E_{\text{exact}} - E_{\text{MCA-MPS}}|$ respectively. The inset displays the ratio of the error of MPS over MCA-MPS, quantifying the improvement achieved by MCA-MPS over standard MPS with same bond dimension. We can find that MCA-MPS can achieve several orders of magnitude and the improvement becomes more dramatic with the increase of bond dimension.}
    \label{Energy_err}
\end{figure}

\begin{figure}[t]
    \centering
    \includegraphics[width=\linewidth]{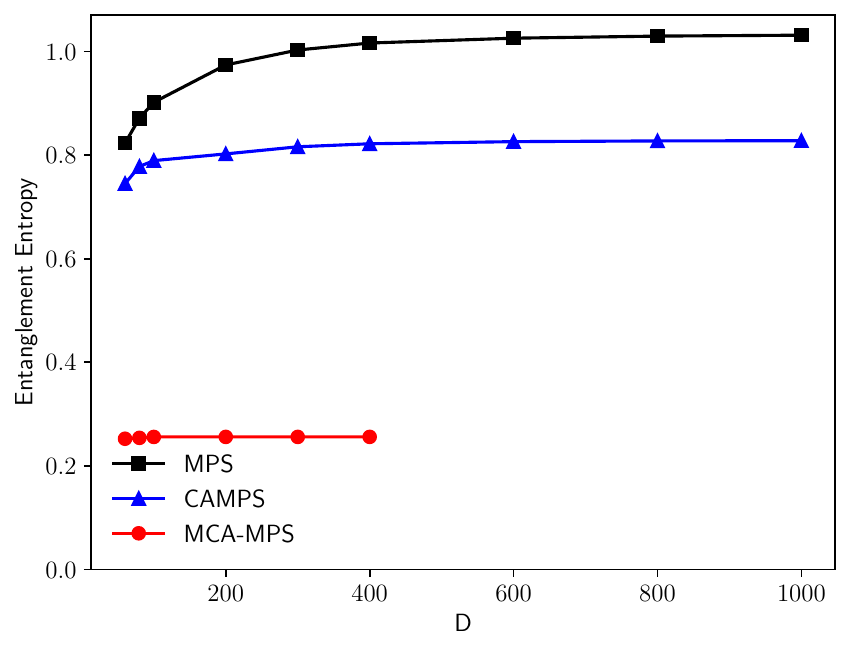}
    \caption{Entanglement entropy of the 1D hydrogen chain \ce{H12} for MPS, CAMPS and MCA-MPS. The entropy is computed for a bipartition of the system into two equal halves and plotted as a function of bond dimension $D$. We can find that the entanglement entropy is converged with small bond dimension in MCA-MPS. The entanglement entropy in MCA-MPS is reduced by a factor about $4$ to the MPS results, consistent with the significant improvement of ground state energy accuracy in Fig.~\ref{Energy_err}. }
    \label{EE}
\end{figure}

\begin{figure}[t]
    \centering
    \includegraphics[width=\linewidth]{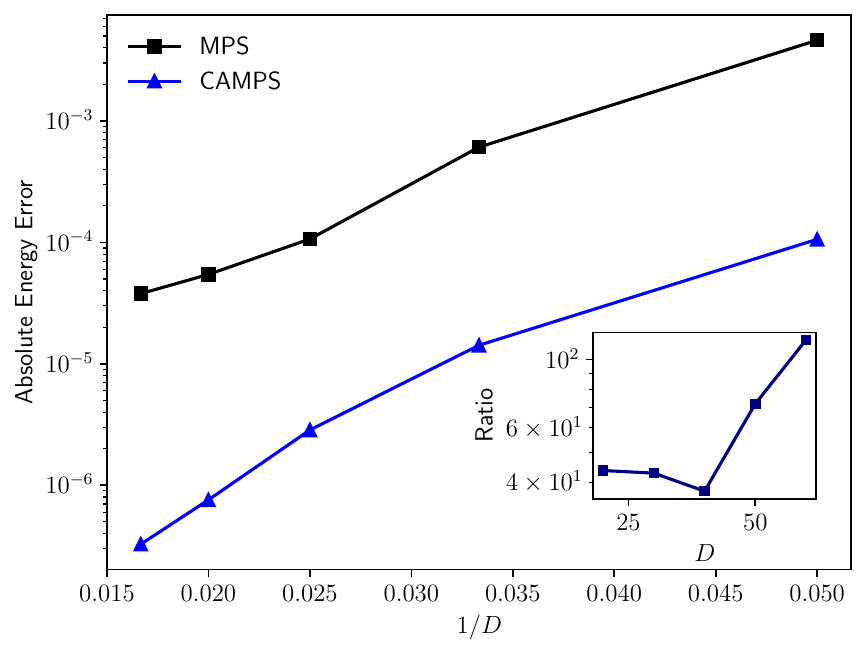}
    \caption{Similar as Fig.~\ref{Energy_err}: energy error (in the unit of Hartree) for the 1D hydrogen chain \ce{H12} relative to exact energy but for OAO orbitals. For OAO orbitals the Matchgate offers no improvement. So only results for MPS and CAMPS are shown.  Errors are defined as $|E_{\text{exact}} - E_{\text{MPS}}|$ and $|E_{\text{exact}} - E_{\text{CAMPS}}|$ respectively. The inset displays the ratio of these errors, quantifying the improvement achieved by CAMPS over standard MPS with same bond dimension. We can find that CAMPS can achieve several orders of magnitude and the improvement becomes more dramatic with the increase of bond dimension.}
    \label{Energy_err_OAO}
    
\end{figure}

In the MCA-MPS framework, $|\text{MPS}\rangle$ is associated with the original Hamiltonian $H$, while the final ansatz $|\text{MCA-MPS}\rangle$ corresponds to the transformed effective Hamiltonian $H_\text{MCA-MPS}$:
\begin{equation}
    H_\text{MCA-MPS} = CGH G^\dagger C^\dagger = \sum_i  a_i P_i
    \label{H_CAMPS}
\end{equation}
where $P_i$ is Pauli string and $a_i$ is the corresponding coefficient. 
We highlight the critical importance of the operational sequence when combining Matchgates and Clifford circuits. We take the common two-body interactions as an example in the following discussion.
While Matchgates can change the number of Hamiltonian's Pauli string composition, they are limited to generating a specific subset of at most $O(N^4)$ (with $N$ the number of orbitals or single particle basis) Pauli strings, because the effect of Matchgate can be represented as $\gamma_i \rightarrow O_{i,j} \gamma_j$ with matrix $O \in SO(2N)$ and $\gamma_i$ are Majorana operators \cite{doi:10.1098/rspa.2008.0189}. Clifford circuits, in contrast, can inter-convert Pauli strings but conserve the Hamiltonian's term count. The optimal strategy involves an initial Matchgate application to transform the Hamiltonian into a structure characterized by $O(N^4)$ terms. Subsequently, Clifford circuits can operate on this structured Hamiltonian, transforming these $O(N^4)$ terms to optimize the ground state representation further. Inverting this order is detrimental: initial Clifford transformations, though preserving the number of Pauli strings, do not generally conserve the interaction form in the Majorana fermion representation. This can convert two-body interactions into complex many-body ones, rendering subsequent Matchgate application, increasing the number of interaction terms in the transformed Hamiltonian.

{\em Model--}
We test the MCA-MPS method in the 1D hydrogen chain \cite{PhysRevX.7.031059} with 12 atoms (\ce{H12}) at an equal bond distance $R=1$ \AA~ in the minimal atomic basis (STO-3G with $N=24$ and the number of total electrons $N_e=12$). We choose the restricted Hartree-Fock (RHF) orbitals for testing our algorithm. The deliberate use of RHF orbitals instead of local atomic basis is to illustrate the power of MCA-MPS. Because the RHF orbitals are delocalized and hence more entangled, which makes the traditional DMRG calculations more challenging. The second quantization Hamiltonian of the system 
for ab initio DMRG\cite{white1999ab,chan2011density,szalay2015tensor,baiardi2020density}
has the following form:
\begin{equation}
    H = \sum_{i,j,\sigma} t_{ij} c^\dagger_{i\sigma} c_{j\sigma} + \sum_{i,j,k,l,\alpha,\beta} V_{ijkl} c^\dagger_{i\alpha} c^\dagger_{j\beta} c_{k\beta} c_{l\alpha},
    \label{Model}
\end{equation}
where $t_{ij}$ and $V_{ijkl}$ are molecular integrals and
$c^{(\dagger)}_{i\sigma}$ are Fermionic annihilation (creation) operators. We only consider particle-number preserved Matchgates (equivalent to orbital transformation) in this work for simplicity. 
In the optimization of Matchgates, we treat each spatial orbital as one site in the DMRG calculation (with local dimension $d =4$ and $D=50$). In the optimization of the Clifford circuits, the spin up and spin down orbitals are treated as two independent sites (with local dimension $d =2$), making the application of Clifford circuit straightforward \cite{2025arXiv250100413H}.    

{\em Results--}
The exact ground-state energy for the hydrogen chain \ce{H12} in the given basis is $-6.452816$ Hartree. In Fig. \ref{Energy_err}, we show the comparison of the error in ground state energy (in the unit of Hartree) with MPS and MCA-MPS. The errors are defined as $|E_{\text{exact}} - E_{\text{MPS}}|$, $|E_{\text{exact}} - E_{\text{CAMPS}}|$ and $|E_{\text{exact}} - E_{\text{MCA-MPS}}|$ respectively. As shown in Fig. \ref{Energy_err}, MCA-MPS method yields substantially more accurate ground-state energies than both MPS and CAMPS method. The error for MCA-MPS is consistently several orders of magnitude smaller. We also notice that the ratio of MPS error to MCA-MPS error grows with bond dimension $D$, as depicted in the inset of Fig. \ref{Energy_err}. This ratio reaches approximately $100$ at $D=100$ and escalates to more than $1000$ by $D=300$, underscoring the powerful synergy of Matchgates and Clifford circuits in enhancing the accuracy of ground-state energy calculations.

Fig.~\ref{EE} illustrates the entanglement entropy (for a mid-chain bipartition) of the 1D \ce{H12} system for MPS, $CAMPS$ and MCA-MPS methods. For MPS, entanglement entropy steadily increases with bond dimension $D$, and for CAMPS it also increases at small $D$. Conversely, the MCA-MPS exhibits a significantly lower entanglement entropy that shows little dependence on $D$. This significant reduction and rapid saturation of entanglement entropy in MCA-MPS suggest that the Matchgates and Clifford circuits efficiently capture the dominant entanglement features, leaving a less entangled state for the subsequent MPS representation. This leads to a more compact and efficient description of the system's overall entanglement.
The dramatic reduction in entanglement entropy is consistent to the significant improvement in the ground state energy accuracy shown in Fig.~\ref{Energy_err}.

We emphasize that the orbital choice critically impacts the performance of the Matchgate and Clifford components in our MCA-MPS ansatz. With orthonormalized atomic orbitals (OAOs), for example, the Matchgate component offers no improvement beyond MPS, whereas the Clifford component still provides a significant error reduction (Fig.~\ref{Energy_err_OAO}). This underscores the necessity of combining both circuit types to ensure optimal performance across general cases (See Supplemental Material \cite{supp} for 2D systems).

As analyzed above and in Supplemental Material \cite{supp}, the sequential optimization reveals a strong synergy. Clifford circuits continue to lower the energy significantly even after Matchgate optimization converges, confirming that they optimize the ground state from different angles. We can also perform a global optimization of Matchgate and Clifford circuits, in which Clifford circuits optimization is applied to the orbitals in the intermediate steps of Matchgate optimizations. This joint optimization method in principle could find global minimum since the converged orbitals from Matchgate optimization isn't necessarily the best orbital for the following Clifford circuits optimizations (See Supplemental Material \cite{supp} for detailed results). 

{\em Discussion--}
In this work, we provide benchmark results of MCA-MPS method on 1D hydrogen chain instead of simpler lattice models because the interaction term in the original Hamiltonian of 1D Hydrogen chain (or general molecule systems) is dense (i.e., scale as $N^4$ with $N$ the number of total orbitals or single particle basis). As was discussed, the transformation of Hamiltonian with particle number preserved Matchgates is equivalent to the unitary transformation of the single particle orbitals, which transform local interactions like the ones in the Hubbard model to long-range ones, significantly increasing the computational complexity even though the transformation with Clifford circuits doesn't increase the number of terms in the Hamiltonian. So to apply MCA-MPS to lattice models like Hubbard model, we need to find ways to reduce the terms in the matchgate-transformed Hamiltonian. In this work, the optimization of matchgates and Clifford circuits are implemented separately. It is also possible to find ways to optimize them simultaneously or iteratively to improve the accuracy. It will be interesting to also implement basis truncation \cite{PhysRevB.81.235129} in MCA-MPS to study systems with large basis set. We only consider U(1) symmetry preserved Matchgates in this work, it is also possible to release this constraint to gain further improvement in certain cases.

{\em Conclusion and perspectives--}
We develop a new wave-function ansatz called MCA-MPS in which MPS is augmented with both Matchgates and Clifford circuits. We also show the optimization of all the MPS, Clifford circuits, and the matchgates in MCA-MPS can be effectively implemented by a minor modification of the DMRG method. Our benchmark results on the one-dimensional hydrogen chain and two dimensional lattice systems show that the MCA-MPS method can significantly improve the accuracy of the ground-state calculation by several orders of magnitude over MPS with the same bond dimension. This new method provides us a useful approach to study quantum many-body systems. The MCA-MPS ansatz also expands our understanding of classically simulatable quantum many-body states. In the future, it will be interesting to explore the combination of other classical many-body methods with different types of classically simulatable quantum circuits. 

{\em Note added:}
Upon completing our work, we became aware of similar works \cite{kang20252dquonlanguageunifying,feng2025quonclassicalsimulationunifying}, where the combination of Clifford circuits, Matchgates and MPS are also discussed. 

\textbf{Acknowledgments:}
MQ acknowledges the support from the National Key Research and Development Program of MOST of China (2022YFA1405400), the National Natural Science Foundation of China (Grant No. 12274290), the Innovation Program for Quantum Science and Technology (2021ZD0301902). 
ZL acknowledges the support from the Quantum Science and Technology-National Science and Technology Major Project (2023ZD0300200) and the Fundamental Research Funds for the Central Universities.

\bibliography{match}

@article{baiardi2020density,
  title={The density matrix renormalization group in chemistry and molecular physics: Recent developments and new challenges},
  author={Baiardi, Alberto and Reiher, Markus},
  journal={J. Chem. Phys.},
  volume={152},
  number={4},
  year={2020},
  publisher={AIP Publishing}
}

@article{white1999ab,
  title={Ab initio quantum chemistry using the density matrix renormalization group},
  author={White, Steven R and Martin, Richard L},
  journal={J. Chem. Phys.},
  volume={110},
  number={9},
  pages={4127--4130},
  year={1999},
  publisher={American Institute of Physics}
}

@article{szalay2015tensor,
  title={Tensor product methods and entanglement optimization for ab initio quantum chemistry},
  author={Szalay, Szil{\'a}rd and Pfeffer, Max and Murg, Valentin and Barcza, Gergely and Verstraete, Frank and Schneider, Reinhold and Legeza, {\"O}rs},
  journal={Int. J. Quantum Chem.},
  volume={115},
  number={19},
  pages={1342--1391},
  year={2015},
  publisher={Wiley Online Library}
}

@article{chan2011density,
  title={The density matrix renormalization group in quantum chemistry},
  author={Chan, Garnet Kin-Lic and Sharma, Sandeep},
  journal={Annu. Rev. Phys. Chem.},
  volume={62},
  number={1},
  pages={465--481},
  year={2011},
  publisher={Annual Reviews}
}

@article{PhysRevB.111.035119,
  title = {Disentangling interacting systems with fermionic Gaussian circuits: Application to quantum impurity models},
  author = {Wu, Ang-Kun and Kloss, Benedikt and Krinitsin, Wladislaw and Fishman, Matthew T. and Pixley, J. H. and Stoudenmire, E. M.},
  journal = {Phys. Rev. B},
  volume = {111},
  issue = {3},
  pages = {035119},
  numpages = {18},
  year = {2025},
  month = {Jan},
  publisher = {American Physical Society},
  doi = {10.1103/PhysRevB.111.035119},
  url = {https://link.aps.org/doi/10.1103/PhysRevB.111.035119}
}

@article{PhysRevX.5.041041,
  title = {{Solutions of the Two-Dimensional Hubbard Model: Benchmarks and Results from a Wide Range of Numerical Algorithms}},
  author = {LeBlanc, J. P. F. and Antipov, Andrey E. and Becca, Federico and Bulik, Ireneusz W. and Chan, Garnet Kin-Lic and Chung, Chia-Min and Deng, Youjin and Ferrero, Michel and Henderson, Thomas M. and Jim\'enez-Hoyos, Carlos A. and Kozik, E. and Liu, Xuan-Wen and Millis, Andrew J. and Prokof\'ev, N. V. and Qin, Mingpu and Scuseria, Gustavo E. and Shi, Hao and Svistunov, B. V. and Tocchio, Luca F. and Tupitsyn, I. S. and White, Steven R. and Zhang, Shiwei and Zheng, Bo-Xiao and Zhu, Zhenyue and Gull, Emanuel},
  collaboration = {Simons Collaboration on the Many-Electron Problem},
  journal = {Phys. Rev. X},
  volume = {5},
  issue = {4},
  pages = {041041},
  numpages = {28},
  year = {2015},
  month = {Dec},
  publisher = {American Physical Society},
  doi = {10.1103/PhysRevX.5.041041},
  url = {https://link.aps.org/doi/10.1103/PhysRevX.5.041041}
}

@ARTICLE{2004cond.mat..7066V,
       author = {{Verstraete}, F. and {Cirac}, J.~I.},
        title = "{{Renormalization algorithms for Quantum-Many Body Systems in two and higher dimensions}}",
      journal = {arXiv e-prints},
         year = 2004,
        month = jul,
          eid = {cond-mat/0407066},
        pages = {cond-mat/0407066},
          doi = {10.48550/arXiv.cond-mat/0407066},
archivePrefix = {arXiv},
       eprint = {cond-mat/0407066},
 primaryClass = {cond-mat.str-el},
       adsurl = {https://ui.adsabs.harvard.edu/abs/2004cond.mat..7066V}
}

@article{PhysRevLett.99.220405,
  title = {{Entanglement Renormalization}},
  author = {Vidal, G.},
  journal = {Phys. Rev. Lett.},
  volume = {99},
  issue = {22},
  pages = {220405},
  numpages = {4},
  year = {2007},
  month = {Nov},
  publisher = {American Physical Society},
  doi = {10.1103/PhysRevLett.99.220405},
  url = {https://link.aps.org/doi/10.1103/PhysRevLett.99.220405}
}

@article{PhysRevLett.102.180406,
  title = {{Entanglement Renormalization in two spatial dimensions}},
  author = {Evenbly, G. and Vidal, G.},
  journal = {Phys. Rev. Lett.},
  volume = {102},
  issue = {18},
  pages = {180406},
  numpages = {4},
  year = {2009},
  month = {May},
  publisher = {American Physical Society},
  doi = {10.1103/PhysRevLett.102.180406},
  url = {https://link.aps.org/doi/10.1103/PhysRevLett.102.180406}
}

@article{PhysRevX.4.011025,
  title = {{Tensor Renormalization of Quantum Many-Body Systems Using Projected Entangled Simplex States}},
  author = {Xie, Z. Y. and Chen, J. and Yu, J. F. and Kong, X. and Normand, B. and Xiang, T.},
  journal = {Phys. Rev. X},
  volume = {4},
  issue = {1},
  pages = {011025},
  numpages = {12},
  year = {2014},
  month = {Feb},
  publisher = {American Physical Society},
  doi = {10.1103/PhysRevX.4.011025},
  url = {https://link.aps.org/doi/10.1103/PhysRevX.4.011025}
}

@article{Qian_2023,
  doi = {10.1088/0256-307X/40/5/057102},
  url = {https://dx.doi.org/10.1088/0256-307X/40/5/057102},
  year = {2023},
  month = {apr},
  publisher = {Chinese Physical Society and IOP Publishing Ltd},
  volume = {40},
  number = {5},
  pages = {057102},
  author = {Xiangjian Qian and Mingpu Qin},
  title = {{Augmenting Density Matrix Renormalization Group with Disentanglers}},
  journal = {Chinese Physics Letters},
}

@article{PhysRevA.70.052328,
  title = {{Improved Simulation of Stabilizer Circuits}},
  author = {Aaronson, Scott and Gottesman, Daniel},
  journal = {Phys. Rev. A},
  volume = {70},
  issue = {5},
  pages = {052328},
  numpages = {14},
  year = {2004},
  month = {Nov},
  publisher = {American Physical Society},
  doi = {10.1103/PhysRevA.70.052328},
  url = {https://link.aps.org/doi/10.1103/PhysRevA.70.052328}
}

@article{PhysRevA.73.022334,
  title = {{Fast Simulation of Stabilizer Circuits using a Graph-state Representation}},
  author = {Anders, Simon and Briegel, Hans J.},
  journal = {Phys. Rev. A},
  volume = {73},
  issue = {2},
  pages = {022334},
  numpages = {9},
  year = {2006},
  month = {Feb},
  publisher = {American Physical Society},
  doi = {10.1103/PhysRevA.73.022334},
  url = {https://link.aps.org/doi/10.1103/PhysRevA.73.022334}
}

@article{PhysRevLett.69.2863,
  title = {{Density matrix formulation for quantum renormalization groups}},
  author = {White, Steven R.},
  journal = {Phys. Rev. Lett.},
  volume = {69},
  issue = {19},
  pages = {2863--2866},
  numpages = {0},
  year = {1992},
  month = {Nov},
  publisher = {American Physical Society},
  doi = {10.1103/PhysRevLett.69.2863},
  url = {https://link.aps.org/doi/10.1103/PhysRevLett.69.2863}
}

@misc{gottesman1997stabilizer,
      title={{Stabilizer Codes and Quantum Error Correction}}, 
      author={Daniel Gottesman},
      year={1997},
      eprint={quant-ph/9705052},
      archivePrefix={arXiv},
      primaryClass={quant-ph}
}

@article{PhysRevLett.75.3537,
  title = {{Thermodynamic limit of Density Matrix Renormalization}},
  author = {\"Ostlund, Stellan and Rommer, Stefan},
  journal = {Phys. Rev. Lett.},
  volume = {75},
  issue = {19},
  pages = {3537--3540},
  numpages = {0},
  year = {1995},
  month = {Nov},
  publisher = {American Physical Society},
  doi = {10.1103/PhysRevLett.75.3537},
  url = {https://link.aps.org/doi/10.1103/PhysRevLett.75.3537}
}

@article{PhysRevLett.128.050402,
  title = {{Stabilizer R\'enyi Entropy}},
  author = {Leone, Lorenzo and Oliviero, Salvatore F. E. and Hamma, Alioscia},
  journal = {Phys. Rev. Lett.},
  volume = {128},
  issue = {5},
  pages = {050402},
  numpages = {5},
  year = {2022},
  month = {Feb},
  publisher = {American Physical Society},
  doi = {10.1103/PhysRevLett.128.050402},
  url = {https://link.aps.org/doi/10.1103/PhysRevLett.128.050402}
}

@article{RevModPhys.93.045003,
  title = {{Matrix product states and projected entangled pair states: Concepts, symmetries, theorems}},
  author = {Cirac, J. Ignacio and P\'erez-Garc\'{\i}a, David and Schuch, Norbert and Verstraete, Frank},
  journal = {Rev. Mod. Phys.},
  volume = {93},
  issue = {4},
  pages = {045003},
  numpages = {65},
  year = {2021},
  month = {Dec},
  publisher = {American Physical Society},
  doi = {10.1103/RevModPhys.93.045003},
  url = {https://link.aps.org/doi/10.1103/RevModPhys.93.045003}
}

@book{Nielsen_Chuang_2010, 
  place={Cambridge}, 
  title={{Quantum Computation and Quantum Information: 10th Anniversary Edition}}, 
  publisher={Cambridge University Press}, 
  author={Nielsen, Michael A. and Chuang, Isaac L.}, 
  year={2010}
}

@book{xiang2023density,
  url={https://books.google.com/books?hl=en&lr=&id=E5fxEAAAQBAJ&oi=fnd&pg=PP1&ots=Hqdq-ApAx0&sig=xi-IvwDkPLk7CTWPcB_d5zRtFZk},
  title={{Density Matrix and Tensor Network Renormalization}},
  author={Xiang, Tao},
  year={2023},
  publisher={Cambridge University Press}
}

@article{doi:10.1098/rspa.2008.0189,
author = {Jozsa, Richard  and Miyake, Akimasa },
title = {Matchgates and classical simulation of quantum circuits},
journal = {Proceedings of the Royal Society A: Mathematical, Physical and Engineering Sciences},
volume = {464},
number = {2100},
pages = {3089-3106},
year = {2008},
doi = {10.1098/rspa.2008.0189},

URL = {https://royalsocietypublishing.org/doi/abs/10.1098/rspa.2008.0189}
}

@article{Veitch_2014,
doi = {10.1088/1367-2630/16/1/013009},
url = {https://dx.doi.org/10.1088/1367-2630/16/1/013009},
year = {2014},
month = {jan},
publisher = {IOP Publishing},
volume = {16},
number = {1},
pages = {013009},
author = {Victor Veitch and S A Hamed Mousavian and Daniel Gottesman and Joseph Emerson},
title = {The resource theory of stabilizer quantum computation},
journal = {New Journal of Physics}
}

@article{PhysRevX.6.021043,
  title = {Trading Classical and Quantum Computational Resources},
  author = {Bravyi, Sergey and Smith, Graeme and Smolin, John A.},
  journal = {Phys. Rev. X},
  volume = {6},
  issue = {2},
  pages = {021043},
  numpages = {14},
  year = {2016},
  month = {Jun},
  publisher = {American Physical Society},
  doi = {10.1103/PhysRevX.6.021043},
  url = {https://link.aps.org/doi/10.1103/PhysRevX.6.021043}
}

@article{RevModPhys.68.13,
  title = {Dynamical mean-field theory of strongly correlated fermion systems and the limit of infinite dimensions},
  author = {Georges, Antoine and Kotliar, Gabriel and Krauth, Werner and Rozenberg, Marcelo J.},
  journal = {Rev. Mod. Phys.},
  volume = {68},
  issue = {1},
  pages = {13--125},
  numpages = {0},
  year = {1996},
  month = {Jan},
  publisher = {American Physical Society},
  doi = {10.1103/RevModPhys.68.13},
  url = {https://link.aps.org/doi/10.1103/RevModPhys.68.13}
}

@article{RevModPhys.78.865,
  title = {Electronic structure calculations with dynamical mean-field theory},
  author = {Kotliar, G. and Savrasov, S. Y. and Haule, K. and Oudovenko, V. S. and Parcollet, O. and Marianetti, C. A.},
  journal = {Rev. Mod. Phys.},
  volume = {78},
  issue = {3},
  pages = {865--951},
  numpages = {0},
  year = {2006},
  month = {Aug},
  publisher = {American Physical Society},
  doi = {10.1103/RevModPhys.78.865},
  url = {https://link.aps.org/doi/10.1103/RevModPhys.78.865}
}

@article{RevModPhys.73.33,
  title = {Quantum Monte Carlo simulations of solids},
  author = {Foulkes, W. M. C. and Mitas, L. and Needs, R. J. and Rajagopal, G.},
  journal = {Rev. Mod. Phys.},
  volume = {73},
  issue = {1},
  pages = {33--83},
  numpages = {0},
  year = {2001},
  month = {Jan},
  publisher = {American Physical Society},
  doi = {10.1103/RevModPhys.73.33},
  url = {https://link.aps.org/doi/10.1103/RevModPhys.73.33}
}

@article{RevModPhys.87.1067,
  title = {Quantum Monte Carlo methods for nuclear physics},
  author = {Carlson, J. and Gandolfi, S. and Pederiva, F. and Pieper, Steven C. and Schiavilla, R. and Schmidt, K. E. and Wiringa, R. B.},
  journal = {Rev. Mod. Phys.},
  volume = {87},
  issue = {3},
  pages = {1067--1118},
  numpages = {52},
  year = {2015},
  month = {Sep},
  publisher = {American Physical Society},
  doi = {10.1103/RevModPhys.87.1067},
  url = {https://link.aps.org/doi/10.1103/RevModPhys.87.1067}
}

@article{PhysRevLett.116.250501,
  title = {Improved Classical Simulation of Quantum Circuits Dominated by Clifford Gates},
  author = {Bravyi, Sergey and Gosset, David},
  journal = {Phys. Rev. Lett.},
  volume = {116},
  issue = {25},
  pages = {250501},
  numpages = {5},
  year = {2016},
  month = {Jun},
  publisher = {American Physical Society},
  doi = {10.1103/PhysRevLett.116.250501},
  url = {https://link.aps.org/doi/10.1103/PhysRevLett.116.250501}
}

@article{PhysRevLett.115.070501,
  title = {Estimating Outcome Probabilities of Quantum Circuits Using Quasiprobabilities},
  author = {Pashayan, Hakop and Wallman, Joel J. and Bartlett, Stephen D.},
  journal = {Phys. Rev. Lett.},
  volume = {115},
  issue = {7},
  pages = {070501},
  numpages = {5},
  year = {2015},
  month = {Aug},
  publisher = {American Physical Society},
  doi = {10.1103/PhysRevLett.115.070501},
  url = {https://link.aps.org/doi/10.1103/PhysRevLett.115.070501}
}

@article{Bravyi2019simulationofquantum,
  doi = {10.22331/q-2019-09-02-181},
  url = {https://doi.org/10.22331/q-2019-09-02-181},
  title = {Simulation of quantum circuits by low-rank stabilizer decompositions},
  author = {Bravyi, Sergey and Browne, Dan and Calpin, Padraic and Campbell, Earl and Gosset, David and Howard, Mark},
  journal = {{Quantum}},
  issn = {2521-327X},
  publisher = {{Verein zur F{\"{o}}rderung des Open Access Publizierens in den Quantenwissenschaften}},
  volume = {3},
  pages = {181},
  month = sep,
  year = {2019}
}

@article{PhysRevX.7.031016,
  title = {Quantum Entanglement Growth under Random Unitary Dynamics},
  author = {Nahum, Adam and Ruhman, Jonathan and Vijay, Sagar and Haah, Jeongwan},
  journal = {Phys. Rev. X},
  volume = {7},
  issue = {3},
  pages = {031016},
  numpages = {30},
  year = {2017},
  month = {Jul},
  publisher = {American Physical Society},
  doi = {10.1103/PhysRevX.7.031016},
  url = {https://link.aps.org/doi/10.1103/PhysRevX.7.031016}
}

@article{PhysRevLett.133.230601,
  title = {Stabilizer Tensor Networks: Universal Quantum Simulator on a Basis of Stabilizer States},
  author = {Masot-Llima, Sergi and Garcia-Saez, Artur},
  journal = {Phys. Rev. Lett.},
  volume = {133},
  issue = {23},
  pages = {230601},
  numpages = {6},
  year = {2024},
  month = {Dec},
  publisher = {American Physical Society},
  doi = {10.1103/PhysRevLett.133.230601},
  url = {https://link.aps.org/doi/10.1103/PhysRevLett.133.230601}
}

@article{RevModPhys.82.277,
  title = {Colloquium: Area laws for the entanglement entropy},
  author = {Eisert, J. and Cramer, M. and Plenio, M. B.},
  journal = {Rev. Mod. Phys.},
  volume = {82},
  issue = {1},
  pages = {277--306},
  numpages = {0},
  year = {2010},
  month = {Feb},
  publisher = {American Physical Society},
  doi = {10.1103/RevModPhys.82.277},
  url = {https://link.aps.org/doi/10.1103/RevModPhys.82.277}
}

@article{PhysRevLett.133.190402,
  title = {Augmenting Density Matrix Renormalization Group with Clifford Circuits},
  author = {Qian, Xiangjian and Huang, Jiale and Qin, Mingpu},
  journal = {Phys. Rev. Lett.},
  volume = {133},
  issue = {19},
  pages = {190402},
  numpages = {6},
  year = {2024},
  month = {Nov},
  publisher = {American Physical Society},
  doi = {10.1103/PhysRevLett.133.190402},
  url = {https://link.aps.org/doi/10.1103/PhysRevLett.133.190402}
}

@article{PhysRevLett.134.150404,
  title = {Clifford Circuits Augmented Time-Dependent Variational Principle},
  author = {Qian, Xiangjian and Huang, Jiale and Qin, Mingpu},
  journal = {Phys. Rev. Lett.},
  volume = {134},
  issue = {15},
  pages = {150404},
  numpages = {6},
  year = {2025},
  month = {Apr},
  publisher = {American Physical Society},
  doi = {10.1103/PhysRevLett.134.150404},
  url = {https://link.aps.org/doi/10.1103/PhysRevLett.134.150404}
}

@ARTICLE{2024arXiv240916895H,
       author = {{Huang}, Jiale and {Qian}, Xiangjian and {Qin}, Mingpu},
        title = "{Non-stabilizerness Entanglement Entropy: a measure of hardness in the classical simulation of quantum many-body systems}",
      journal = {arXiv e-prints},
         year = 2024,
        month = sep,
          eid = {arXiv:2409.16895},
        pages = {arXiv:2409.16895},
          doi = {10.48550/arXiv.2409.16895},
archivePrefix = {arXiv},
       eprint = {2409.16895},
 primaryClass = {quant-ph},
       adsurl = {https://ui.adsabs.harvard.edu/abs/2024arXiv240916895H}
}

@ARTICLE{2024arXiv241015709Q,
       author = {{Qian}, Xiangjian and {Huang}, Jiale and {Qin}, Mingpu},
        title = "{Augmenting Finite Temperature Tensor Network with Clifford Circuits}",
      journal = {arXiv e-prints},
         year = 2024,
        month = oct,
          eid = {arXiv:2410.15709},
        pages = {arXiv:2410.15709},
          doi = {10.48550/arXiv.2410.15709},
archivePrefix = {arXiv},
       eprint = {2410.15709},
 primaryClass = {quant-ph},
       adsurl = {https://ui.adsabs.harvard.edu/abs/2024arXiv241015709Q}
}

@ARTICLE{2025arXiv250100413H,
       author = {{Huang}, Jiale and {Qian}, Xiangjian and {Qin}, Mingpu},
        title = "{Clifford circuits Augmented Matrix Product States for fermion systems}",
      journal = {arXiv e-prints},
         year = 2024,
        month = dec,
          eid = {arXiv:2501.00413},
        pages = {arXiv:2501.00413},
          doi = {10.48550/arXiv.2501.00413},
archivePrefix = {arXiv},
       eprint = {2501.00413},
 primaryClass = {cond-mat.str-el},
       adsurl = {https://ui.adsabs.harvard.edu/abs/2025arXiv250100413H}
}

@article{PhysRevB.111.085121,
  title = {Disentangling critical quantum spin chains with Clifford circuits},
  author = {Fan, Chaohui and Qian, Xiangjian and Zhang, Hua-Chen and Huang, Rui-Zhen and Qin, Mingpu and Xiang, Tao},
  journal = {Phys. Rev. B},
  volume = {111},
  issue = {8},
  pages = {085121},
  numpages = {8},
  year = {2025},
  month = {Feb},
  publisher = {American Physical Society},
  doi = {10.1103/PhysRevB.111.085121},
  url = {https://link.aps.org/doi/10.1103/PhysRevB.111.085121}
}

@article{PRXQuantum.6.010345,
  title = {Quantum State Designs with Clifford-Enhanced Matrix Product States},
  author = {Lami, Guglielmo and Haug, Tobias and De Nardis, Jacopo},
  journal = {PRX Quantum},
  volume = {6},
  issue = {1},
  pages = {010345},
  numpages = {14},
  year = {2025},
  month = {Mar},
  publisher = {American Physical Society},
  doi = {10.1103/PRXQuantum.6.010345},
  url = {https://link.aps.org/doi/10.1103/PRXQuantum.6.010345}
}

@inproceedings{10.1145/380752.380785,
author = {Valiant, Leslie G.},
title = {Quantum computers that can be simulated classically in polynomial time},
year = {2001},
isbn = {1581133499},
publisher = {Association for Computing Machinery},
address = {New York, NY, USA},
url = {https://doi.org/10.1145/380752.380785},
doi = {10.1145/380752.380785},
booktitle = {Proceedings of the Thirty-Third Annual ACM Symposium on Theory of Computing},
pages = {114–123},
numpages = {10},
location = {Hersonissos, Greece},
series = {STOC '01}
}

@ARTICLE{2024arXiv241205367C,
       author = {{Collura}, Mario and {De Nardis}, Jacopo and {Alba}, Vincenzo and {Lami}, Guglielmo},
        title = "{The quantum magic of fermionic Gaussian states}",
      journal = {arXiv e-prints},
         year = 2024,
        month = dec,
          eid = {arXiv:2412.05367},
        pages = {arXiv:2412.05367},
          doi = {10.48550/arXiv.2412.05367},
archivePrefix = {arXiv},
       eprint = {2412.05367},
 primaryClass = {quant-ph},
       adsurl = {https://ui.adsabs.harvard.edu/abs/2024arXiv241205367C}
}

@ARTICLE{2024arXiv241010068P,
       author = {{Projansky}, Andrew M. and {Necaise}, Jason and {Whitfield}, James D.},
        title = "{Gaussianity and Simulability of Cliffords and Matchgates}",
      journal = {arXiv e-prints},
         year = 2024,
        month = oct,
          eid = {arXiv:2410.10068},
        pages = {arXiv:2410.10068},
          doi = {10.48550/arXiv.2410.10068},
archivePrefix = {arXiv},
       eprint = {2410.10068},
 primaryClass = {quant-ph},
       adsurl = {https://ui.adsabs.harvard.edu/abs/2024arXiv241010068P}
}

@article{PhysRevB.81.235129,
  title = {Optimizing Hartree-Fock orbitals by the density-matrix renormalization group},
  author = {Luo, H.-G. and Qin, M.-P. and Xiang, T.},
  journal = {Phys. Rev. B},
  volume = {81},
  issue = {23},
  pages = {235129},
  numpages = {4},
  year = {2010},
  month = {Jun},
  publisher = {American Physical Society},
  doi = {10.1103/PhysRevB.81.235129},
  url = {https://link.aps.org/doi/10.1103/PhysRevB.81.235129}
}

@article{PhysRevLett.117.210402,
  title = {Fermionic Orbital Optimization in Tensor Network States},
  author = {Krumnow, C. and Veis, L. and Legeza, \"O. and Eisert, J.},
  journal = {Phys. Rev. Lett.},
  volume = {117},
  issue = {21},
  pages = {210402},
  numpages = {6},
  year = {2016},
  month = {Nov},
  publisher = {American Physical Society},
  doi = {10.1103/PhysRevLett.117.210402},
  url = {https://link.aps.org/doi/10.1103/PhysRevLett.117.210402}
}

@misc{kang20252dquonlanguageunifying,
      title={2D Quon Language: Unifying Framework for Cliffords, Matchgates, and Beyond}, 
      author={Byungmin Kang and Chen Zhao and Zhengwei Liu and Xun Gao and Soonwon Choi},
      year={2025},
      eprint={2505.06336},
      archivePrefix={arXiv},
      primaryClass={quant-ph},
      url={https://arxiv.org/abs/2505.06336}, 
}

@article{PhysRevX.7.031059,
  title = {Towards the Solution of the Many-Electron Problem in Real Materials: Equation of State of the Hydrogen Chain with State-of-the-Art Many-Body Methods},
  author = {Motta, Mario and Ceperley, David M. and Chan, Garnet Kin-Lic and Gomez, John A. and Gull, Emanuel and Guo, Sheng and Jim\'enez-Hoyos, Carlos A. and Lan, Tran Nguyen and Li, Jia and Ma, Fengjie and Millis, Andrew J. and Prokof'ev, Nikolay V. and Ray, Ushnish and Scuseria, Gustavo E. and Sorella, Sandro and Stoudenmire, Edwin M. and Sun, Qiming and Tupitsyn, Igor S. and White, Steven R. and Zgid, Dominika and Zhang, Shiwei},
  collaboration = {Simons Collaboration on the Many-Electron Problem},
  journal = {Phys. Rev. X},
  volume = {7},
  issue = {3},
  pages = {031059},
  numpages = {28},
  year = {2017},
  month = {Sep},
  publisher = {American Physical Society},
  doi = {10.1103/PhysRevX.7.031059},
  url = {https://link.aps.org/doi/10.1103/PhysRevX.7.031059}
}

@misc{feng2025quonclassicalsimulationunifying,
      title={Quon Classical Simulation: Unifying Clifford, Matchgates and Entanglement}, 
      author={Zixuan Feng and Zhengwei Liu and Fan Lu and Ningfeng Wang},
      year={2025},
      eprint={2505.07804},
      archivePrefix={arXiv},
      primaryClass={quant-ph},
      url={https://arxiv.org/abs/2505.07804}, 
}

@ARTICLE{2024arXiv241111720F,
       author = {{Frau}, Martina and {Sonya Tarabunga}, Poetri and {Collura}, Mario and {Tirrito}, Emanuele and {Dalmonte}, Marcello},
        title = "{Stabilizer disentangling of conformal field theories}",
      journal = {arXiv e-prints},
         year = 2024,
        month = nov,
          eid = {arXiv:2411.11720},
        pages = {arXiv:2411.11720},
          doi = {10.48550/arXiv.2411.11720},
archivePrefix = {arXiv},
       eprint = {2411.11720},
 primaryClass = {quant-ph},
       adsurl = {https://ui.adsabs.harvard.edu/abs/2024arXiv241111720F}
}

@article{PhysRevLett.134.150403,
  title = {Clifford Dressed Time-Dependent Variational Principle},
  author = {Mello, Antonio Francesco and Santini, Alessandro and Lami, Guglielmo and De Nardis, Jacopo and Collura, Mario},
  journal = {Phys. Rev. Lett.},
  volume = {134},
  issue = {15},
  pages = {150403},
  numpages = {7},
  year = {2025},
  month = {Apr},
  publisher = {American Physical Society},
  doi = {10.1103/PhysRevLett.134.150403},
  url = {https://link.aps.org/doi/10.1103/PhysRevLett.134.150403}
}

@ARTICLE{li2025entanglementminimizedorbitalsenablefaster,
       author = {{Li}, Zhendong},
        title = "{Entanglement-minimized orbitals enable faster quantum simulation of molecules}",
      journal = {arXiv e-prints},
         year = 2025,
        month = jun,
          eid = {arXiv:2506.13386},
        pages = {arXiv:2506.13386},
          doi = {10.48550/arXiv.2506.13386},
archivePrefix = {arXiv},
       eprint = {2506.13386},
 primaryClass = {quant-ph},
       adsurl = {https://ui.adsabs.harvard.edu/abs/2025arXiv250613386L}
}

@misc{supp,
    note={See Supplemental Material for additional details about the results of 2D hydrogen lattice systems benchmark, the numerical results of synergy between Matchgate and Clifford Circuits and the optimization of Matchgate}
}

\clearpage
\onecolumngrid
\begin{center}
\textbf{\large Supplementary Materials for ``Augmenting Density Matrix Renormalization Group with Matchgates and Clifford circuits''}
\end{center}
\vspace{1cm}
\twocolumngrid

\appendix
\setcounter{equation}{0}
\setcounter{figure}{0}
\setcounter{table}{0}
\setcounter{page}{1}
\setcounter{section}{0}
\renewcommand{\theequation}{S\arabic{equation}}
\renewcommand{\thefigure}{S\arabic{figure}}
\renewcommand{\thetable}{S\arabic{table}}

\tableofcontents

\section{2D results}
\label{2D_results}

    

\begin{figure}[t]
    \centering
    \includegraphics[width=\linewidth]{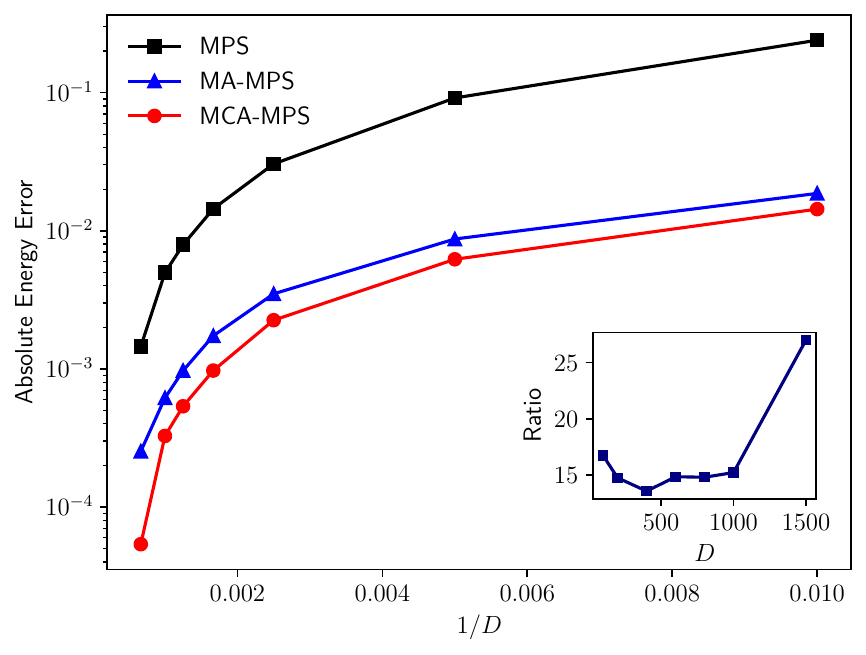}
    \caption{2D OAO results for MPS, MA-MPS and MCA-MPS.
Errors (in the unit of Hartree) are defined as  $|E_{\text{ref}} - E_{\text{MPS}}|$, $|E_{\text{ref}} - E_{\text{MA-MPS}}|$ and $|E_{\text{ref}} - E_{\text{MCA-MPS}}|$ respectively.
The inset displays the ratio of MPS error to MCA-MPS error, quantifying the improvement achieved by MCA-MPS over standard MPS with same bond dimension.
}
    \label{Energy_err_OAO_2D}
\end{figure}

\begin{figure}[h]
    \centering
    \includegraphics[width=\linewidth]{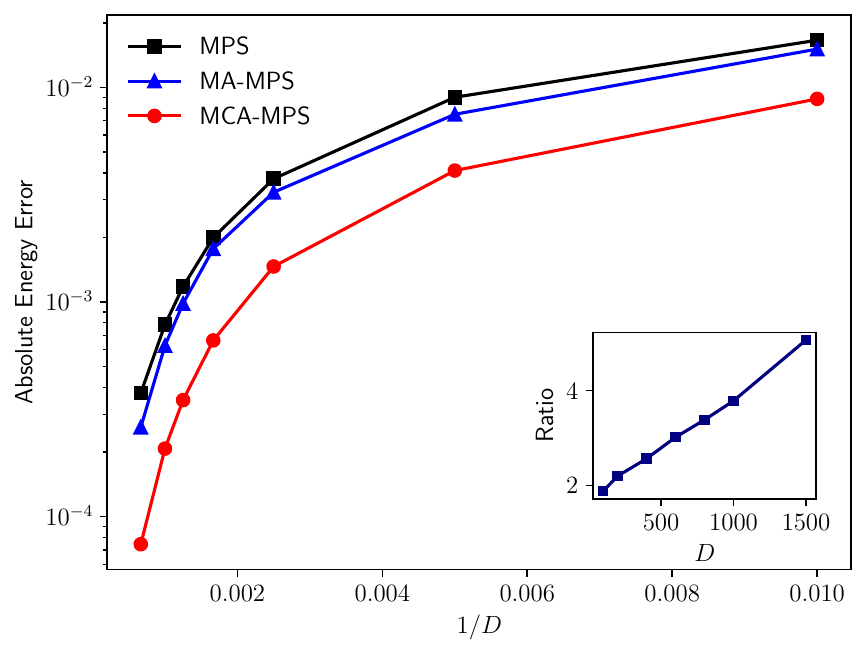}
    \caption{2D LMO results for MPS, MA-MPS and MCA-MPS.
Errors (in the unit of Hartree) are defined as  $|E_{\text{ref}} - E_{\text{MPS}}|$, $|E_{\text{ref}} - E_{\text{MA-MPS}}|$ and $|E_{\text{ref}} - E_{\text{MCA-MPS}}|$ respectively.
The inset displays the ratio of MPS error to MCA-MPS error, quantifying the improvement achieved by MCA-MPS over standard MPS with same bond dimension.}
    \label{Energy_err_LMO_2D}
\end{figure}

To test our method on 2D systems, we simulate a 16-atom ($4$ by $4$) hydrogen plane ($R=1$ \AA) in a minimal basis, corresponding to $N=32$ orbitals at half-filling ($N_e=16$).
We compare the performance of standard MPS, MA-MPS (Matchgate-augmented MPS), and our full MCA-MPS ansatz.
These methods are benchmarked across two orbital basis (orthonormalized atomic orbitals (OAOs) and localized molecular orbitals (LMOs)) against a high-accuracy SU(2) DMRG reference energy of $E_{\text{ref}} = -7.785104$ ($D=3000$).
The detailed results for each orbital basis are presented in Figs.~\ref{Energy_err_OAO_2D}-\ref{Energy_err_LMO_2D}.

While MCA-MPS consistently outperforms MPS, the relative contributions of the Matchgate and Clifford components are strongly basis-dependent.
For OAO orbitals, both components contribute significantly, leading to the most pronounced improvement in MCA-MPS.
The results from LMO basis are different, where Matchgates offer a modest gain and Clifford circuits provide a significant one.

This basis-dependent performance highlights the complementary roles of Matchgates and Clifford circuits, likely in capturing different aspects of the electronic correlation.
It thus emphasizes the importance of their combined use within the MCA-MPS framework to achieve robust performance across diverse orbital representations.

\section{Joint optimization of Matchgate and Clifford circuits}
\label{Matchgate_Clifford}
A potential concern with sequential optimization is the risk of trapping the ansatz in a local minimum, which would preclude access to the global ground state.
We demonstrate here that the two components act synergistically rather than competitively.

We also perform a joint optimization of Matchgate and Clifford circuits in which Clifford circuit optimization is applied to each intermediate orbitals in the Matchgate optimization process.

\begin{figure}[t]
    \centering
    \includegraphics[width=\linewidth]{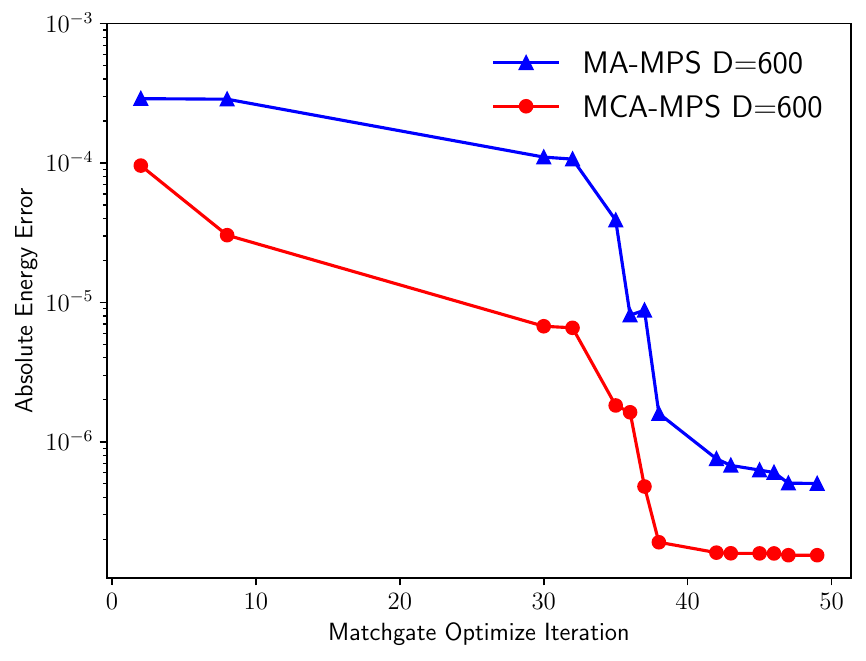}
    \caption{Evolution of the ground-state energy error (in the unit of Hartree) for the \ce{H12} chain.
The red curve represents the baseline MA-MPS ($D=600$) optimized solely via Matchgates.
The other curves (MCA-MPS) show energies obtained by applying Clifford optimization to intermediate MA-MPS states (starting from $D=100$) at iteration steps.}
    \label{fig:Matchgate_Iteration}
\end{figure}

Fig.~\ref{fig:Matchgate_Iteration} tracks the energy evolution of the \ce{H12} chain with restricted Hartree-Fock orbital.
The MA-MPS ansatz (blue curve, $D=600$) serves as a baseline, reflecting energy minimization driven exclusively by Matchgate updates.
To assess the full MCA-MPS performance, we extract system snapshots at various Matchgate iterations and subject them to subsequent Clifford circuit optimization.

The Matchgate optimization approaches convergence after approximately 40 iterations.
Crucially, the full MCA-MPS ansatz consistently outperforms the MA-MPS baseline throughout the entire process.
Even in the saturation regime (iterations 40--50), the introduction of Clifford circuits yields a substantial energy reduction, lowering the energy error from $\sim 5 \times 10^{-7}$ to $1.5 \times 10^{-7}$.
If the Matchgate and Clifford sectors were strongly competitive, the optimal Matchgate circuits for a pure MA-MPS would differ drastically from that required for the full MCA-MPS ansatz, rendering the sequential procedure inefficient.
Instead, the consistent improvement implies that these components target different Hilbert space sectors.
Matchgates minimize Gaussian entanglement, while Clifford circuits resolve the remaining non-stabilizerness.
Consequently, this decoupling allows the sequential optimization to effectively approach the global ground state.

In principle, the joint optimization could find a global minimum since the converged orbitals in the Matchgate optimization process is not necessarily the best orbitals for Clifford circuit optimization.
But in this specific case, it turns out the converged orbitals from the Matchgate optimization indeed gives the best orbitals for the following Clifford circuit optimization.
Nevertheless, in practical calculation, the joint optimization should be tested to find global minimum in MCA-MPS.

\section{The optimization of Matchgate}
\label{Matchgate_opt}
To optimize the matchgate, in this work we adopted the randomized orbital optimization algorithm
in Ref.
\cite{li2025entanglementminimizedorbitalsenablefaster}, which simultaneously optimize
the energy and reduce the entanglement. Specifically, for a given MPS, the givens rotations between adjacent sites are applied to minimize the entanglement
\begin{equation}
(\tilde{\hat{a}}_{i\sigma}^\dagger,\tilde{\hat{a}}_{j\sigma}^\dagger)
=
(\hat{a}_{i\sigma}^\dagger,\hat{a}_{j\sigma}^\dagger)
\mathbf{G}(\theta),\;\;
\mathbf{G}(\theta)=
\begin{bmatrix}
\cos\theta & -\sin\theta \\
\sin\theta & \cos\theta
\end{bmatrix}.
\label{eq:givens}
\end{equation}
Then, a layer of random swaps ($\theta\in\{0,\pi/2\}$) in a linear layout is applied
in order to jump out of local minimum, followed by a sequence of givens rotations to prevent the random swaps creating too large perturbation.
The obtained set of gates are applied to transform the Hamiltonian,
and a small number of DMRG sweeps is applied to minimize the energy.
Based on the optimized
energy and entanglement entropy, a decision is made to accept or reject the current set of gates.
This procedure is repeated for a number of iterations until the energy lowering is small enough.

To elucidate the quantum circuit implementation of the orbital optimization, we establish the equivalence between the givens rotation in Eq.~(\ref{eq:givens}) and the nearest-neighbor matchgates.
The linear transformation of operators corresponds to a unitary gate $\hat{U}(\theta)$ acting on the two-site Hilbert space spanned by $\{|00\rangle, |01\rangle, |10\rangle, |11\rangle\}$.
In this basis, the vacuum $|00\rangle$ is invariant under the passive rotation.
The single-particle sector transforms according to $\mathbf{G}(\theta)$, while the doubly occupied state transforms as $\hat{U}|11\rangle = (\det\mathbf{G})|11\rangle$.
Since $\det\mathbf{G}=1$ for the Givens rotation, the $|11\rangle$ state remains unchanged.
This yields the explicit matrix representation of a particle-conserving matchgate:
\begin{equation}
\mathcal{M}(\theta) =
\begin{bmatrix}
1 & 0 & 0 & 0 \\
0 & \cos\theta & -\sin\theta & 0 \\
0 & \sin\theta & \cos\theta & 0 \\
0 & 0 & 0 & 1
\end{bmatrix}.
\label{eq:matchgate}
\end{equation}

\end{document}